\documentclass[preprint]{JHEP3}

\usepackage{epsfig,multicol}

\title{Holographic QCD beyond the leading order}
\author{
        Youngman Kim,
                ~P. Ko \\ 
        School of Physics, Korea Institute for Advanced Study,
         207-43, Cheongryangri 2-dong, Dongdaemun-gu, Seoul 130-722, Korea
}
\author{
        Xiao-Hong Wu$^{1,2}$ \\
        $^1$ Institute of Modern Physics, School of Science,
        East China University of Science and Technology,
        Meilong Road 130, Shanghai 200237, China \\
        $^2$ School of Physics, Korea Institute for Advanced Study,
         207-43, Cheongryangri 2-dong, Dongdaemun-gu, Seoul 130-722, Korea
}
\received{}
\revised{}
\accepted{}

\preprint{arXiv:0804.2710 [hep-ph]}

\abstract{
We consider a holographic QCD model for light mesons beyond the leading
order in the context of 5-dim gauged linear sigma model on the interval
in the AdS$_5$ space. We include two dimension-6 operators in addition to
the canonical bulk kinetic terms, and study chiral dynamics of $\pi$,
$\rho$, $a_1$ and some of their KK modes.
As novel features of dim-6 operators, we get non-vanishing
Br$(a_1 \to \pi \gamma)$, the electromagnetic form factor and
the charge radius of a charged pion, which improve the leading order
results significantly and agree well with the experimental results.
}


\begin{document}

\section{Introduction}

To understand the dynamics of low lying hadrons from underlying QCD
has been a long standing problem in theoretical physics.
In chiral Lagrangian approaches, it has  been known for some time that
the low energy dynamics of pions, vector mesons $\rho$ and axial vector
mesons $a_1$ are well described by the gauged linear sigma model
(or its nonlinear version) with massive Yang-Mills gauge filds.
The model Lagrangian up to  dimension-6 operators is given by
\footnote{We ignore the Wess-Zumino-Witten term in this work.}
\begin{eqnarray}
{\cal L}_{\rm Massive YM} & = & {\rm Tr}
 \left[ -{1\over 4} L_{\mu\nu} L^{\mu\nu}
- {1\over 4} R_{\mu\nu} R^{\mu\nu} + {1\over 2} D_\mu \Phi D^\mu \Phi
- {1\over 2} M_\Phi^2 \Phi^{\dagger} \Phi \right]
\nonumber  \\
& + & {1\over 2} m_0^2~{\rm Tr} ( L_\mu L^\mu + R_\mu R^\mu )
\nonumber  \\
& + & {\rm Tr} \left[
               + \zeta \left( L_{\mu\nu} D^\mu \Phi D^\nu \Phi^{\dagger}
                  + R_{\mu\nu} D^\mu \Phi^{\dagger} D^\nu \Phi \right)
               + \kappa L_{\mu\nu} \Phi R^{\mu\nu} \Phi^{\dagger} \right]
\nonumber  \\
& + & \lambda_1 {\rm Tr} ( \Phi^{\dagger} \Phi )^2
  +   \lambda_2 \left[ {\rm Tr} ( \Phi^{\dagger} \Phi ) \right]^2
  +   ( \lambda_3 {\rm det} ( \Phi ) + H.c. )
\end{eqnarray}
The role of the higher dimensional operators in light hadron dynamics,
especially $\kappa$ and $\xi$ terms, were studied  in the framework of
the gauged linear sigma model in 4D~\cite{Meissner:1987ge, Ko:1994en}.
Although the above Lagrangian is quite successful in describing the
$\pi-\rho-a_1$ system, it has a conceptual drawback in that we need to
give gauge boson masses $m_0^2$ by hand.
If we put $m_0^2 = 0$, global chiral symmetry becomes local symmetry,
which is not a true symmetry of real QCD, and we end up with massless
$\rho$ and $a_1$, which is phenomenologically disastrous.
Therefore we have to put $m_0^2 \neq 0$ and have to impose chiral
symmetry only as a global symmetry. However, if chiral symmetry is only
a global symmetry, then there is no compelling reason to introduce
gauge covariant derivative, hence no reason for the minimal coupling
between hadrons and (axial) vector mesons, and thus universality of the
$P-V-V$ couplings. Since the universality seems to hold to a good
approximation, it is tempting to implement global chiral symmetry to
local symmetry. This have been remained a problem in chiral dynamics
approach to the low lying hadrons.

Recently, there have been many interesting and successful attempts
to understand hadron physics in the context of AdS/CFT correspondence
~\cite{AdSCFT}.
The properties of hardrons and the hadron physics phenomenology are
studied in various approaches ~\cite{TDold, SS056, EKSS,  DaRold:2005zs,
  Hirn:2005vk, Brodsky, ET, Yee, swwx, h2rs, RecenthQCD}
which are inspired by the AdS/CFT correspondence.

One may start with some stringy setup that may reproduce certain
aspects of nonperturbative QCD.
The most successful approach so far seems, arguably, the works by Sakai
and Sugkimoto \cite{SS056}, and follow-up papers \cite{topdownz}.
The model by Sakai and Sugimoto has nice features, but also some drawbacks.
They show that chiral dynamics of $\pi$, $\rho$ and $a_1$ system can be
well reproduced by studying the $N_f$ D8-branes  in the background of
$N_c$ D4-branes. Also the Wess-Zumino-Witten (WZW) term is derived from
the 5-dim Chern-Simon (CS) term. On the other hand, there are spurious
$SO(5)$ symmetry from $S^5$, which is not a true symmetry of real QCD.
And it is not easy to accommodate nonzero quark masses, namely nonzero
pion mass. Finally the pion and its radial excitation comes from different
5-dim fields, which is not easy to understand within the quark model.
Despite numerous remarkable successes of Sakai-Sugimito model, there is
an ample room for further improvement.

Independent of the stringy approach, a gravity dual model of the gauged
linear sigma model was proposed to describe the chiral dynamics of light
hadrons~\cite{EKSS, DaRold:2005zs}. This approach is often called the bottom-up approach,
where one starts from QCD and then tries to construct
its five-dimensional holographic dual model, AdS/QCD.
Following the AdS/CFT correspondence, it is assumed that there are bulk
fields that couple to the 4-dimensional QCD operators.
For example, there are bulk gauge fields $L_M$ and  $R_M$ that couple
to the QCD operators $j_L \equiv \bar{q_L} \gamma^\mu q_L$ and
$j_R \equiv \bar{q}_R \gamma^\mu q_R$, which are flavor currents.

Quality of the overall fit to the meson properties in the models
of Ref.~\cite{EKSS, DaRold:2005zs} is at the level of $\sim 30 \%$,
which is quite remarkable, considering the simplicity of the model.
However it predicts $B( a_1 \rightarrow \pi \gamma ) = 0$ and
too small charge radius of a charged pion.
The Lagrangian in  Ref.~\cite{EKSS, DaRold:2005zs} is the leading order
one, since it contains only the bulk kinetic terms for the bulk gauge
fields and scalar fields. In order to improve the predictions for
$B ( a_1 \rightarrow \pi \gamma )$ and the charge radius of a charged
pion, we have to go beyond the leading order Lagrangian.

In this paper, we construct an AdS$_5$ dual model of the
gauged linear sigma model with dimension-6 operators, motivated by
the recently developed AdS/QCD model~\cite{EKSS, DaRold:2005zs}.
To this end we incorporate higher dimensional operators, especially
two dim-6 terms, into the AdS/QCD model.
\footnote{For a brief report on the present work, see Ref~\cite{Wu:2006zz}}
In this work, we only consider the vector, axial-vector and pseudoscalar
sectors, as a first step of our study.
Interestingly enough, we find that the aforementioned problem of
giving gauge boson masses $m_0^2$ is no longer present, since one can give
masses of the vector and axial vector mesons, by projecting out the zero
modes by choosing suitable boundary conditions. Degeneracy between the
vector and the axial vector mesons will be lifted by the conventional
Higgs mechanism. Still there remain physical pions.

Naively, these new operators will
have nontrivial effects on the interaction vertex,
as well as mass spectra and decay constants.
We expect that they may contribute to the Br$(a_1 \to \pi \gamma)$,
which is zero in the original
AdS/QCD model~\cite{EKSS, DaRold:2005zs}.
We also study the phenomenology of
$\rho \to \pi \pi$ and $a_1 \to \rho \pi$,
the branching ratios and
D/S wave amplitude ratio in the latter channel.
By introducing photon as an external field,
we study the pion electromagnetic form factor,
and calculate the pion electromagnetic charge radius,
which agrees with the experimental results in our numerical study.

This paper is organized as follows.
In section 2, we define the Lagrangian of our model with
two dim-6 operators in AdS$_5$.
In section 3, we study the mass spectra and decay constants in
vector, axial-vector and pseudoscalar sectors.
We also present the interaction vertex and
phenomenology of $a_1 \to \rho \pi$,
$\rho \to \pi \pi$, $a_1 \to \pi \gamma$ channels,
and calculate the pion charge radius.
We derive the relevant chiral coefficients in section 4,
and give our numerical results in section 5.
The conclusions are drawn in section 6.

\section{Gauged Linear Sigma Model in the AdS$_5$ space}

The Lagrangian of the holographic QCD model~\cite{EKSS, DaRold:2005zs}
defined in a slice of AdS$_5$ is given by
\begin{eqnarray}
{\cal L}_{\rm 5}^{\rm dim-4} & = & \sqrt{g}
M_5 ~{\rm Tr}~\left[ -{1\over 4} L_{MN} L^{MN}
- {1\over 4} R_{MN} R^{MN} \right.
\nonumber  \\
& + & \left.  {1\over 2} (D_M \Phi)^\dagger D^M \Phi
- {1\over 2} M_\Phi^2 \Phi^{\dagger} \Phi \right]\, ,\label{hQCDL4}
\end{eqnarray}
where $M_\Phi^2 = -3/L^2$ from AdS/CFT correspondence
~\cite{AdSCFT}, $D_M\Phi =\partial_M\Phi +iL_M\Phi -i\Phi R_M$,
$L_M=L_M^a \tau^a/2$ with $\tau^a$ being the Pauli matrix,
and $M, N = 0,1,2,3,5 ({\rm or}\,z\,)$.
We define $\Phi = S e^{i P/v(z)}$ with $\langle S \rangle = v(z)$.
 Under ${\rm SU(2)}_V$, $S$ and $P$ transform as singlet and
 triplet, respectively.
The AdS$_5$ space is  characterized in the conformally
flat metric with a warp factor $a(z) \equiv L/z$:
\begin{equation}
ds^2 =  a^2 (z) ( dx^\mu dx_\mu - dz^2 ).
\end{equation}
The scale $L$ is the curvature of the 5-dimensional AdS space.
 In this model, the AdS$_5$ space is compactified such that
$L_0<z<L_1$, where $L_0\rightarrow 0$ is an ultra-violet (UV) cutoff
 and $L_1$ is an infrared (IR) cutoff.
Solving the equation of motion for $S$, we obtain~\cite{DaRold:2005zs}
\begin{eqnarray}
\langle S \rangle \equiv v(z) = c_1 z + c_2 z^3
\end{eqnarray}
with the integration constants $c_{1,2}$,
\begin{eqnarray}
c_1 = \frac{M_q L_1^3 - \xi L_0^2}{L L_1 (L_1^2 - L_0^2)},
 \qquad c_2 = \frac{\xi - M_q L_1}{L L_1 (L_1^2 - L_0^2)}\; .
\end{eqnarray}
Here we adopted the following boundary conditions
\begin{eqnarray}
M_q = \frac{L}{L_0} v \bigg|_{L_0}, \qquad \xi = L v \bigg|_{L_1},
\end{eqnarray}
where $M_q$ is the current quark mass matrix, which
breaks chiral symmetry explicitly,
and $\xi$ is related to $\langle \bar{q} q \rangle$,
which  breaks chiral symmetry spontaneously.
 The value of $L_1$ is fixed by
the rho-meson mass: $1/L_1\simeq 320~ {\rm MeV}$~\cite{EKSS,
  DaRold:2005zs}. There may be several ways to improve the model given
above, though several observables obtained from the model
are in agreement with experiments. One immediate extension of the
model is to see corrections from various sources:
trilinear or quartic interactions among the vector
fields, 5D loop corrections, higher dimensional operators
and back-reactions on the metric due to condensates~\cite{swwx}.
In the present work, we consider corrections to the model from higher
dimensional operators, though to be consistent we have to treat all those
corrections at the same time.
We note that a part of large $N_c$ corrections  through
meson-loop contributions are discussed in Ref.~\cite{Harada:2006di}.

Now we introduce higher dimensional operators in the model
Lagrangian in Eq. (\ref{hQCDL4}). In principle, we can include
infinite tower of higher dimensional operators, but for simplicity
we consider only dimension-6 operators in the chiral limit.
Note here that we have the following mass dimensions
for a scalar field $\Phi$ and vector fields $L_M$ and $R_M$:
\begin{equation}
{\rm dim } ( \Phi ) = {\rm dim} ( L_M ) = {\rm dim} ( R_M ) = 1\, .
\end{equation}
The Lagrangian with dimension-6 operators reads
\begin{eqnarray}
{\cal L}_{\rm 5}^{\rm dim-6}  & = & \sqrt{g}M_5 ~{\rm Tr} \bigg[
      - i \frac{\kappa}{M_5^2} \bigg( L_{MN} D^M \Phi (D^N \Phi)^{\dagger}
        + R_{MN} (D^M \Phi)^{\dagger} D^N \Phi \bigg) \nonumber\\
&&\qquad\quad + \frac{\zeta}{M_5^2} L_{MN} \Phi R^{MN} \Phi^{\dagger}
\bigg] \, ,\label{hQCDL6}
\end{eqnarray}
where $\kappa$ and $\zeta$ are constants that will be fixed later.

There are more dimension--6 operators, such as
\begin{equation}
{\cal L}_{\rm 5}^{\rm dim-6}  = \sqrt{g}M_5 ~{\rm Tr} \bigg[
L_M^{~N}  L_N^{~P} L_P^{~M} + ( L \leftrightarrow R )
\bigg]
\end{equation}
However these terms are $O( p^6 )$ after chiral symmetry breaking,
whereas the $\kappa$ and $\zeta$ terms are $O( p^4 )$ after chiral
symmetry breaking. Therefore we keep only those dimension--6 terms
that reduce to $O ( p^4 )$ after chiral symmetry breaking.
We note that the corrections to physical observables
from the second dim-6 operator Tr$[L_{MN} \Phi R^{MN} \Phi^{\dagger}]$
in Eq. (2.7) and the operator in Eq. (2.8) have been discussed
in Ref.~\cite{Grigoryan:2007iy}.

\section{Vector, axial-vector and pseudoscalar sectors}

\subsection{Relevant parts of the Lagrangian}

In this section, we work in the chiral limit.
Then $v(z)$ is proportional to $\mathbf{1}$,
$v(z) \simeq \xi \frac{z^3}{L_1^3} \mathbf{1}$.
The vector and axial gauge bosons are defined by
\begin{eqnarray}
V_M &=& \frac{1}{\sqrt{2}} (L_M + R_M) \nonumber\\
A_M &=& \frac{1}{\sqrt{2}} (L_M - R_M)\, .
\end{eqnarray}
In order to cancel the mixing terms of
$V_{\mu}$, $A_{\mu}$ ($\mu$ as 4D Lorentz index, $0,1,2,3$)
and $V_z$, $A_z$, $P$,
we add gauge fixing terms
\begin{eqnarray}
{\cal L}^V_{\rm GF} &=& - \frac{M_5 a}{2 \xi_V} {\rm Tr}
  \bigg[ \partial_\mu V^\mu - \frac{\xi_V}{a} \bigg( \partial_5 (a V_z)
         - \frac{2 \zeta}{M_5^2} \partial_5 (a v^2 V_z) \bigg) \bigg]^2
  \, ,\nonumber\\
{\cal L}^A_{\rm GF} &=& - \frac{M_5 a}{2 \xi_A} {\rm Tr}
  \bigg[ \partial_\mu A^\mu - \frac{\xi_A}{a} \bigg( \partial_5 (a A_z)
         + \sqrt{2} a^3 v P \nonumber\\
&&       + \frac{2 \sqrt{2} \kappa}{M_5^2} \partial_5 (a(\partial_5 v) P)
         + \frac{4 \kappa}{M_5^2} a v (\partial_5 v) A_z
         + \frac{2 \zeta}{M_5^2} \partial_5 (a v^2 A_z) \bigg)
  \bigg]^2\, .
\end{eqnarray}
In the unitary gauge, $\xi_{V,A} \to \infty$, we have the following relation
between $A_z$ and $P$,
\begin{eqnarray}
\label{azprelation}
\sqrt{2} a^3 v P + \partial_5 (a A_5)
         + \frac{2 \sqrt{2} \kappa}{M_5^2} \partial_5 (a(\partial_5 v) P)
         + \frac{4 \kappa}{M_5^2} a v (\partial_5 v) A_z
         + \frac{2 \zeta}{M_5^2} \partial_5 (a v^2 A_z) = 0,
\end{eqnarray}
which is identical to the leading order relation~\cite{DaRold:2005zs}
when $\kappa=\zeta=0$.

The quadratic terms for vector, axial-vector and pseudoscalar are given by,
after integration by parts,
\begin{eqnarray}
{\cal L}_V &=& \frac{M_5}{2} a {\rm Tr} \bigg\{
  V_\mu \bigg( \partial^2 Z_v
  - a^{-1} \partial_5 a Z_v \partial_5  \bigg) V^\mu \bigg\}\, , \nonumber\\
{\cal L}_A &=& \frac{M_5}{2} a {\rm Tr} \bigg\{
  A_\mu \bigg( \partial^2 Z_a
  - a^{-1} \partial_5 a Z_a \partial_5 + 2 a^2 v^2
  - \frac{8\kappa}{M_5^2} v (\partial_5 v) \partial_5  \bigg) A^\mu
  \bigg\}\, , \nonumber\\
{\cal L}_\pi &=& \frac{M_5}{2} a {\rm Tr} \bigg\{
  (- 2 a^3 v^2) \bigg( A_z + \partial_5 \frac{P}{\sqrt{2} v} \bigg) \nonumber\\
&&  + a (\partial_\mu A_z)^2 + a^3 (\partial_\mu P)^2
  + \frac{4 \sqrt{2} \kappa}{M_5^2} a (\partial_5 v)
   (\partial_\mu A_z) (\partial^\mu P )
  + \frac{2 \zeta}{M_5^2} a v^2 (\partial_\mu A_z)^2 \bigg\}\, .
\label{quardratic}
\end{eqnarray}
with $Z_v = 1 - \frac{2 \zeta v^2}{M_5^2}$ and
$Z_a = 1 + \frac{2 \zeta v^2}{M_5^2}$.
The boundary terms are
\begin{eqnarray}
{\cal L}_{\rm boundary} &=& M_5 a {\rm Tr} \bigg(
  V^\mu Z_v \partial_5 V_\mu
  + A^\mu Z_a \partial_5 A_\mu \nonumber\\
&&\qquad  - A_\mu \partial^\mu A_z
  - \frac{2 \sqrt{2} \kappa}{M_5^2} (\partial_5 v) A_\mu \partial^\mu P
  - \frac{2 \zeta}{M_5^2} v^2 A_\mu \partial^\mu A_z \bigg)
  \bigg|_{z=L_0}^{z=L_1}\, .
\end{eqnarray}
We choose the following boundary conditions to cancel the IR-boundary terms,
\begin{eqnarray}
\label{irboundary}
&& \partial_5 V_\mu\bigg|_{z=L_1} = \partial_5 A_\mu\bigg|_{z=L_1} = 0, \quad
V_5\bigg|_{z=L_1} = A_5\bigg|_{z=L_1} = 0 \\
\label{irboundarypion}
&&A_z + \frac{2 \sqrt{2} \kappa}{M_5^2} (\partial_5 v) P
  + \frac{2 \zeta}{M_5^2} v^2 A_z\bigg|_{z=L_1} = 0\, ,
\end{eqnarray}
and the UV-boundary condition will be specified later.

We also calculated $VAP$, $VPP$ and four-pion interaction vertices
\begin{eqnarray}
{\cal L}_{VAP} &=& \frac{\sqrt{2} i}{2} M_5 a {\rm Tr} \bigg[
  A^\mu [\partial_5 V_\mu, A_z] - (\partial_5 A^\mu) [V_\mu, A_z]
  + \sqrt{2} a^2 v A^\mu [V_\mu, P] \bigg] \nonumber\\
&& - \frac{2 i \kappa}{M_5} a {\rm Tr} \bigg[
  (\partial_5 v) (\partial_5 A^\mu) [V_\mu, P]
   - \sqrt{2} v (\partial_5 v) A^\mu [V_\mu, A_z]
   + \sqrt{2} v^2 A^\mu [\partial_5 V_\mu, A_z] \nonumber\\
&&\qquad\qquad   + v A^\mu [\partial_5 V_\mu, \partial_5 P]
   + v A^\mu [V_{\mu\nu}, \partial^\nu P] \bigg] \nonumber\\
&& - \frac{i \zeta}{M_5} a v {\rm Tr} \bigg[
   A^{\mu\nu} [V_{\mu\nu}, P] - 2 (\partial_5 A^\mu) [\partial_5 V_\mu, P]
   + \sqrt{2} v A^\mu [\partial_5 V_\mu, A_z] \nonumber\\
&&\qquad\qquad   + \sqrt{2} v (\partial_5 A^\mu) [V_\mu, A_z] \bigg]
  \, ,
\label{VAP}\\
{\cal L}_{V\pi\pi} &=& \frac{i}{2} M_5 a {\rm Tr} \bigg[
  V^\mu [A_z, \partial_\mu A_z] + a^2 V^\mu [P, \partial_\mu P] \bigg]
   \nonumber\\
&& +\frac{\sqrt{2}i\kappa}{M_5} a \bigg[\frac{1}{2}
   V^{\mu\nu} [\partial_\mu P, \partial_\nu P]
   + (\partial_5 V^\mu) [\partial_\mu P, \partial_5 P]
   - \sqrt{2} v (\partial_5 V^\mu) [A_z, \partial_\mu P] \nonumber\\
&&\qquad\qquad   + \sqrt{2} (\partial_5 v) V^\mu [A_z, \partial_\mu P]
   - \sqrt{2} (\partial_5 v) V^\mu [\partial_\mu A_z, P] \bigg] \nonumber\\
&& - \frac{\sqrt{2}i\zeta}{M_5} a v {\rm Tr} \bigg[
   - v V^\mu [A_z, \partial_\mu A_z]
   + \sqrt{2} (\partial_5 V^\mu) [P, \partial_\mu A_z] \bigg] \, ,
\label{VPP}\\
{\cal L}_{\pi^4} &=& - \frac{a^3 M_5}{12 v^2} {\rm Tr} \bigg[
   (\partial_\mu P)^2 P^2 - \bigg( (\partial_\mu P) P \bigg)^2 \bigg]
   \nonumber\\
&& - \frac{a\kappa}{M_5} {\rm Tr} \bigg[
   (\partial^\mu A_z) (\partial_\mu P) A_z P
    - \bigg( (\partial_\mu A_z) P \bigg)^2 \bigg] \nonumber\\
&& + \frac{\sqrt{2}}{3} \frac{a (\partial_5 v)\kappa}{v^2 M_5} {\rm Tr} \bigg[
   (\partial^\mu A_z) (\partial_\mu P) P P
    - (\partial^\mu A_z) P (\partial_\mu P) P \bigg] \nonumber\\
&& - \frac{\sqrt{2} a\kappa}{v M_5} {\rm Tr} \bigg[
   (\partial^\mu A_z) (\partial_\mu P) P (\partial_5 P)
    - (\partial^\mu A_z) (\partial_5 P) (\partial_\mu P) P \bigg] \nonumber\\
&& - \frac{a \zeta}{M_5} {\rm Tr} \bigg[
   (\partial_\mu A_z)^2 P^2 - \bigg( (\partial_\mu A_z) P \bigg)^2
   \bigg]\, .
\label{pi4}
\end{eqnarray}

\subsection{Two-point correlation functions}

We calculate the two-point correlation functions for vector and
axial-vector
with respect to the UV boundary external source fields $v_\mu$ and $a_\mu$,
which couple to the vector and axial-vector currents operators, respectively,
\begin{eqnarray}
\label{vasource}
V_\mu|_{z=L_0} = v_\mu, \qquad A_\mu|_{z=L_0} = a_\mu.
\end{eqnarray}
From the AdS/CFT correspondence,
in order to calculate the current-current correlation function
in the strongly coupled CFT side,
we can do it in the weakly interacting AdS side instead.
Then the effective Lagrangian in momentum space
in term of the correlators is
\begin{eqnarray}
{\cal L}_{\rm eff} &=& v_\mu \Pi^{\mu\nu}_V(p^2) v_\nu
  + a_\mu \Pi^{\mu\nu}_A(p^2) a_\nu
\end{eqnarray}
with $\Pi^{\mu\nu}_{V,A}(p^2) =
 (g^{\mu\nu} - p^\mu p^\nu/p^2) \, \Pi_{V,A}(p^2)$.
We solve the equations of motion for the vector and axial-vector field
derived from Eq.(\ref{quardratic})
with the boundary conditions (\ref{irboundary}) and (\ref{vasource})
and calculate the two-point correlation function
\begin{eqnarray}
\Pi(p^2) &=& - M_5 L \frac{\partial_5 f(z)}{z f(z)} \bigg|_{z=L_0\to 0}
\label{corr}
\end{eqnarray}
where $f(z)$ is the solution of differential equation.
With dim-6 operators, we cannot calculate
the two-point correlation function $\Pi(p^2)$ analytically,
instead, we do it numerically.

For asymptotically large momentum $p^2 L_1^2 \gg 1$,
we can expand the 2-point functions in powers of $1/p^2$, and get
\begin{eqnarray}
\Pi_{V,A}(p^2) &=& p^2 \bigg[
  \frac{M_5 L}{2} \ln p^2 L_0^2
  + c_6^{V,A} \frac{1}{p^6}  
  \bigg],
\end{eqnarray}
with
\begin{eqnarray}
c_6^V &=& - \frac{192 \zeta}{5 M_5 L L_1^6} \xi^2,
\qquad c_6^A = \bigg( \frac{16 M_5 L}{5 L_1^6}
   + \frac{192 \zeta}{5 M_5 L L_1^6}
   + \frac{384 \kappa}{5 M_5 L L_1^6} \bigg) \xi^2,
\end{eqnarray}
which agree with the results in Ref.~\cite{DaRold:2005zs}
for $\kappa = \zeta = 0$.
It is worthwhile to calculate the left-right correlator
$\Pi_{LR} = \Pi_V - \Pi_A$, in the large momentum limit, we have
\begin{eqnarray}
\Pi_{LR} = \frac{c_6}{p^4} + ... \,
\end{eqnarray}
with $c_6 = c_V - c_A$,
where the experiment value $c_6 = - 4\pi \alpha_s \langle \bar{q} q \rangle^2
\simeq - 1.3 \times 10^{-3} {\rm GeV}^6$ obtained from Ref.~\cite{Jamin}.
We remark here that the vector correlator obtained in the present work and
in the hard wall model ~\cite{EKSS, DaRold:2005zs}
has no $1/p^4$ compared to the results from
operator product expansion (OPE)~\cite{Shifman}. 
In the chiral limit,
the coefficient of $1/p^4$ term is due to the gluon condensate~\cite{Shifman}.
In the hard wall model adopted in the present work, however,
the metric is just a pure AdS with no gluon condensate included,
and the model has no 5D bulk scalar field that couples to
tr$(G_{\mu\nu}G^{\mu\nu})$ at the boundary,
 where $G_{\mu\nu}$ is the gluon field strength tensor.
Therefore, the vector and axial-vector correlators in the hard wall model
do not contain $1/p^4$ term, as it should be.
To have $1/p^4$ in Eq. (3.14), we have to consider
a deformed AdS background~\cite{gluoncond}
due to the back-reaction of the gluon condensate.


In the large N$_c$ limit, the above correlators can be written as the sum
in terms of the resonance masses and decay constants,
\begin{eqnarray}
\Pi_A(p^2) &=& p^2 \sum_n \frac{f^2_{A_n}}{p^2 - M_{A_n}^2} + f_\pi^2 \\
\Pi_V(p^2) &=& p^2 \sum_n \frac{f^2_{V_n}}{p^2 - M_{V_n}^2} .
\end{eqnarray}
Then the vector and axial meson masses are determined as
the poles of their corresponding correlators,
and the decay constants are related with the residue,
\begin{eqnarray}
f^2_{\rho, a_1} &=& \lim_{p^2 \to m^2_{\rho, a_1}}
  (p^2 - m^2_{\rho, a_1}) \Pi_{V,A}(p^2)/p^2 , \\
f^2_\pi &=& \Pi_A(0) .
\end{eqnarray}

\section{Interactions and phenomenology}

\subsection{ KK decompositions}

Now we study hadronic observables such as decay widths
and form factors using our model given in Eq.~ (\ref{hQCDL4}) and
Eq.~ (\ref{hQCDL6}). Primarily we investigate how those dimension-6
operators in Eq.~ (\ref{hQCDL6}) affect the results obtained with only
interactions in  Eq.~ (\ref{hQCDL4}).
To this end,
we first Kaluza-Klein (KK) decompose the vector field as
$V_\mu(x,z) = \frac{1}{\sqrt{M_5 L}} \sum_{n=1}^{\infty}
 \tilde{V}^{(n)}_\mu(x) f^{(n)}_V(z)$
and also for the axial-vector and pseudoscalar fields,
where we omit the superscript index $(n)$ when we consider the lowest KK mode.
The first resonances of the vector, axial-vector and pseudoscalar fields
are associated with $\rho$, $a_1$ and $\pi$ respectively.
The equations of motion for the vector, axial-vector and
pseudoscalar fields are easily read off
from Eq.~(\ref{quardratic}). To cancel the boundary terms,
in addition to the IR boundary conditions given in Eq.~(\ref{irboundary}) and
Eq. (\ref{irboundarypion}), we impose the following UV boundary conditions
\begin{eqnarray}
V_\mu \bigg|_{z=L_0} = 0, \qquad A_\mu \bigg|_{z=L_0} = 0,
\qquad P \bigg|_{z=L_0} = 0\, .
\end{eqnarray}
We obtain the wave function and mass spectra of various fields numerically
with the normalization conditions:
\begin{eqnarray}
\label{normalization}
&& \int_{L_0}^{L_1} dz \frac{a}{L} Z_v (z) f^{(m)}_V(z)  f^{(n)}_V(z) =
\delta_{mn},
  \nonumber\\
&& \int_{L_0}^{L_1} dz \frac{a}{L} Z_a (z) f^{(m)}_A(z) f^{(n)}_A(z)  =
\delta_{mn},
  \nonumber\\
&& \int_{L_0}^{L_1} dz \frac{a}{L} \bigg( (f_{A_z}(z))^2 + a^2 (f_P(z))^2
  + \frac{4\sqrt{2}\kappa}{M_5^2} (\partial_5 v) f_{A_z} f_P
  + \frac{2\zeta}{M_5^2} v^2 (f_{A_z})^2 \bigg) = 1.
\end{eqnarray}

\subsection{ $\rho  \rightarrow \pi \pi$}
The $\rho \pi \pi$ vertex can be expressed as
\begin{eqnarray}
{\cal L}_{\rho\pi\pi} = \frac{i}{\sqrt{2}} g_{\rho\pi\pi}
   {\rm Tr} (\tilde{V}^\mu [\tilde{A}_5, \partial_\mu \tilde{A}_5])
  + \frac{i}{\sqrt{2}} f_{\rho\pi\pi}
   {\rm Tr} (\tilde{V}^{\mu\nu} [\partial_\mu \tilde{A}_5, \partial_\nu
    \tilde{A}_5])
\end{eqnarray}
with the couplings
\begin{eqnarray}
g_{\rho\pi\pi} &=& \int_{L_0}^{L_1}dz \frac{a}{\sqrt{M_5 L^3}} \bigg[
  f_V f_{A_z}^2 + a^2 f_V f_P^2 \nonumber\\
&&\qquad  + \frac{2\kappa}{M_5^2} \bigg( - (\partial_5 f_V) f_P (\partial_5 f_P)
   - \sqrt{2} v (\partial_5 f_V) f_{A_z} f_P
   + 2 \sqrt{2} (\partial_5 v) f_V f_{A_z} f_P \bigg) \nonumber\\
&&\qquad  - \frac{2\zeta v}{M_5^2} \bigg( - v f_V f_{A_z}^2
   + \sqrt{2} (\partial_5 f_V) f_{A_z} f_P \bigg) \bigg] \\
f_{\rho\pi\pi} &=& \int_{L_0}^{L_1}dz
   \frac{\kappa}{\sqrt{M_5^5 L^3}} a f_V f_P^2
\end{eqnarray}
We also calculate the decay width $\Gamma(\rho \to \pi \pi)$,
which includes the non-minimal coupling $f_{\rho\pi\pi}$,
even though its contribution is numerically small.

\subsection{Electromagnetic form factor of a charged pion}

Before we study the electromagnetic form factor of a charged pion,
we introduce the photon as an external gauge field and rewrite
the bulk vector field decomposition as
\begin{eqnarray}
\label{vectorkk}
V_\mu(x,z) = e \tilde{F}_\mu(x) \tau_3
  + \frac{1}{\sqrt{M_5 L}} \sum_{n=1}^{\infty}
   \tilde{V}^{(n)}_\mu(x) f^{(n)}_V(z)\, ,
\end{eqnarray}
with $\tau_3 = \sigma_3/\sqrt{2}$, where $\sigma$ is the Pauli matrix,
and $e$ is identified with the physical electron charge
at chiral symmetry breaking scale.
To treat photon and $\rho$ on the same footing,
we introduce $f_F(z) = 1$ as the fifth dimension profile for photon.
The advantage of our treatment of photon as external field,
compared with the treatment of photon as the electromagnetic subgroup
of $SU(3)_V$~\cite{DaRold:2005zs},
is that we don't need to worry about the KK excitations of the photon,
as well as the mixing between photon KK excitations and $\rho^0$
KK excitations.

We consider the electromagnetic form factors of pions.
In additional to the usual structure of contact $\gamma \pi \pi$ interaction
${\rm Tr} (F^\mu [\tilde{A}_5, \partial_\mu \tilde{A}_5])$,
we also have non-minimal structure
${\rm Tr} (F^{\mu\nu} [\partial_\mu \tilde{A}_5, \partial_\nu \tilde{A}_5])$,
which comes from the dim-6 $\kappa$ term.
And we also find $g_{\gamma \pi\pi} = e$ after comparing with
the pion normalization condition, eq.(\ref{normalization}).
From the kinetic term (\ref{quardratic})
and the vector KK decomposition (\ref{vectorkk}),
we can derive the kinetic mixing of $\gamma$ and $\rho$,
\begin{eqnarray}
{\cal L}_{\gamma\rho} = - \frac{1}{2} e g_{\gamma\rho} F^{\mu\nu}
\tilde{V}_{\mu\nu}\, ,
\end{eqnarray}
with
\begin{eqnarray}
g_{\gamma\rho} = \frac{M_5}{\sqrt{M_5 L}} \int_{L_0}^{L_1} dz a Z_v
f_V(z)\, .
\end{eqnarray}
The electromagnetic form factor of pion can be calculated as
\begin{eqnarray}
F(q^2) &=& 1 - \frac{f_{\gamma\pi\pi}}{g_{\gamma\pi\pi}} q^2 -
\frac{g_{\gamma\rho} q^2}{q^2 - m_\rho^2} g_{\rho\pi\pi} .
\end{eqnarray}
In small momentum limit, it can also be expressed as
\begin{eqnarray}
F(q^2) = 1 + \frac{1}{6} r_\pi^2 q^2 + {\cal O}(q^4)\, ,
\end{eqnarray}
with the pion charge radius $r_\pi$ calculated as
\begin{eqnarray}
r_\pi^2 = 6 \bigg[ - \frac{f_{\gamma\pi\pi}}{g_{\gamma\pi\pi}}
  + \frac{g_{\gamma\rho} g_{\rho\pi\pi}}{m_\rho^2} \bigg]\, .
\end{eqnarray}
Our vector meson dominance (VMD) is different from
the usual VMDs as discussed in ref.~\cite{O'Connell:1995wf},
where we have an additional non-minimal $\gamma \pi \pi$ contact interaction,
${\rm Tr} (\tilde{F}^{\mu\nu}
[\partial_\mu \tilde{A}_5, \partial_\nu \tilde{A}_5])$.

\subsection{$a_1 \to \rho \pi$} 
We first consider the process $a_1 \to \rho \pi$.
Applying the KK-decomposition to ${\cal L}_{VAP}$ in Eq. (\ref{VAP}), we obtain
\begin{eqnarray}
{\cal L}_{a_1\rho\pi} &=& i g_{1 a_1\rho\pi}
   {\rm Tr} (\tilde{A}^\mu [\tilde{V}_\mu, \tilde{A}_z])
  + i g_{2 a_1\rho\pi}
   {\rm Tr} (\tilde{A}^\mu [\tilde{V}_{\mu\nu}, \partial^\nu \tilde{A}_z])
  \nonumber\\
&& + i g_{3 a_1\rho\pi}
   {\rm Tr} (\tilde{A}^{\mu\nu} [\tilde{V}_{\mu\nu}, \tilde{A}_z])
\end{eqnarray}
with the coefficients $g_{ia_1\rho\pi}$ ($i=1,2,3$)
\begin{eqnarray}
g_{1 a_1\rho\pi} &=& \int_{L_0}^{L_1}dz \frac{a}{\sqrt{M_5 L^3}} \bigg[
  \frac{1}{\sqrt{2}} \bigg( f_A (\partial_5 f_V) f_{A_z}
   - (\partial_5 f_A) f_V f_{A_z}
   + \sqrt{2} a^2 v f_A f_V f_P \bigg) \nonumber\\
&&\quad - \frac{2 \kappa}{M_5^2} \bigg( (\partial_5 v) (\partial_5 f_A) f_V f_P
   - \sqrt{2} v (\partial_5 v) f_A f_V f_{A_z} \nonumber\\
&&\quad   + \sqrt{2} a v^2 f_A (\partial_5 f_V) f_{A_z}
   + a v f_A (\partial_5 f_V) (\partial_5 f_P) \bigg) \nonumber\\
&&\quad  - \frac{\sqrt{2}\zeta}{M_5^2} a v \bigg(
   v (\partial_5 f_A) f_V f_{A_z}
   + v f_A (\partial_5 f_V) f_{A_z}
   - \sqrt{2} (\partial_5 f_A) (\partial_5 f_V) f_P \bigg) \bigg] \\
g_{2 a_1\rho\pi} &=& - \int_{L_0}^{L_1}dz
   \frac{2 \kappa}{\sqrt{M_5^5 L^3}} \bigg[
   a v f_A f_V f_P \bigg] \\
g_{3 a_1\rho\pi} &=& - \int_{L_0}^{L_1}dz
   \frac{\sqrt{2}\zeta}{\sqrt{M_5^5 L^3}} \bigg[
   a v^2 (\partial_5 f_A) f_V f_{A_z} \bigg]\, .
\end{eqnarray}
With the interaction vertex above, it is straightforward
to derive the amplitude of the process,
which can be written as
\begin{eqnarray}
{\cal A}(a_1 \to \rho \pi) = - i \epsilon^\mu(s_{a_1})
  \epsilon^\nu(s_\rho) \bigg[
  f_{a_1\rho\pi} g_{\mu\nu} + g_{a_1\rho\pi} {p_\pi}_\mu {p_\pi}_\nu
  \bigg] \, .\nonumber
\end{eqnarray}
The S/D wave amplitudes are defined as in Ref.~\cite{Isgur:1988vm}
\begin{eqnarray}
\langle \rho(\vec{k} s_\rho) \pi(-\vec{k}) | H | a_1(0s_{a_1}) \rangle &=&
  i f^S_{a_1\rho\pi} \delta_{s_\rho s_{a_1}} Y_{00}(\Omega_k)
  + i f^D_{a_1\rho\pi} \sum_{m_L} C(211; m_L s_\rho s_{a_1})
  Y_{2m_L}(\Omega_k)\, ,
 \nonumber
\end{eqnarray}
with
\begin{eqnarray}
f^S_{a_1\rho\pi} &=& \frac{\sqrt{4\pi}}{3 m_\rho} \bigg[
  (E_\rho + 2 m_\rho) f_{a_1\rho\pi} - k^2 m_{a_1} g_{a_1\rho\pi} \bigg]
  \nonumber\\
f^D_{a_1\rho\pi} &=& - \frac{\sqrt{8\pi}}{3 m_\rho} \bigg[
  (E_\rho - m_\rho) f_{a_1\rho\pi} - k^2 m_{a_1} g_{a_1\rho\pi}
  \bigg]\, .
\end{eqnarray}
And also the decay width of $a_1 \to \rho \pi$ is
\begin{eqnarray}
\Gamma(a_1 \to \rho \pi) &=& \frac{p_c}{4\pi m_{a_1}} \bigg[
  \frac{2}{3} f_{a_1\rho\pi}^2 + \frac{1}{3}
  \bigg( \frac{E_\rho}{m_\rho} f_{a_1\rho\pi}
   + \frac{m_{a_1}}{m_\rho} p_c^2 g_{a_1\rho\pi} \bigg)^2 \bigg]\, .
\end{eqnarray}

\subsection{$a_1 \to \pi \gamma$} 
In this subsection, we study the process $a_1 \to \pi \gamma$,
With the help of vector KK decomposition Eq.(\ref{vectorkk}),
we have similar results as $a_1 \to \rho \pi$.
We have verified that the gauge non-invariant term of structure
${\rm Tr}(\tilde{A}^\mu [F_\mu, \tilde{A}_z])$ is cancelled out,
when we impose the relation between $A_z$ and $P$,
e.g., Eq.~(\ref{azprelation}),
the boundary condition Eq.~(\ref{irboundarypion}) and $\partial_5 f_F=0$.

\subsection{Numerical results}

\begin{figure}[!htb]
\centerline{\includegraphics[width=8cm] {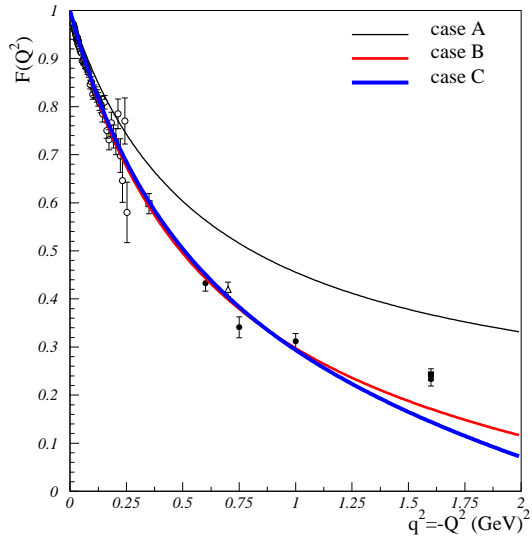}}
\caption{ \label{pionformfactor}
Pion form factor $F(Q^2)$ as a function of $q^2$.
The white circles are data from CERN~\cite{Amendolia:1983di},
square from DESY~\cite{Brauel:1977ra},
triangle from DESY~\cite{Ackermann:1977rp},
black circle from Jlab~\cite{Tadevosyan:2007yd},
and black square from Jlab~\cite{Horn:2006tm}.
}
\end{figure}

In this subsection, we present the numerical results of various hadronic
obsevables and chiral coefficients discussed previously.
We use $\chi^2$ to fit the four parameters $L_1$, $\xi$, $\kappa$, $\zeta$
from $m_\rho$, $m_{a_1}$, D$/$S ratio, $\Gamma(\rho \to \pi \pi)$ in case B
and $m_\rho$, $m_{a_1}$, $\Gamma(\rho \to \pi \pi)$, $f_\pi$ in case C.
Our results are summarized in Table~\ref{table1},~\ref{tableope},
and Figure ~\ref{pionformfactor}.
As a comparison, we also give
Da Rold and Pomarol's results~\cite{DaRold:2005zs} in case A.

In both cases B and C,
$\Gamma(a_1 \to \pi \gamma)$ is non-vanishing, but small (less than $100$KeV),
while $\Gamma(a_1 \to \rho \pi)$ is a little small in case B,
but consistent with experimental measurement in case C.
We have checked that the dominant contribution to $\Gamma(a_1 \to \rho \pi)$
comes from the leading order structure
${\rm Tr} (\tilde{A}^\mu [\tilde{V}_\mu, \tilde{A}_z])$.
However, ${\rm Tr} (\tilde{A}^\mu [\tilde{F}_\mu, \tilde{A}_z])$ term
is not gauge invariant and cancelled out for $a_1 \to \pi \gamma$ channel.
Then only dim-6 $\kappa$ and $\zeta$ terms contribute to the above process.
This is different from usual 4D models with large $\Gamma(a_1 \to \pi \gamma)$,
where the ratio between $\Gamma(a_1 \to \pi \gamma)$ and
$\Gamma(a_1 \to \rho \pi)$ is roughly $e^2/g_{\rho\pi\pi}^2$,
and only a single type of operator
${\rm Tr} (\tilde{A}^{\mu\nu} [\tilde{V}_{\mu\nu}, \pi])$
contributes to both channels~\cite{Meissner:1987ge}.

\begin{table}[!htb]
\small{
\begin{center}
\begin{tabular}{|c|c|c|c|c|c|c|c|}
\hline
case & $L_1$ & $\kappa$ ($10^{-6}$) & $m_{\rho}$ & $m_{a_1}$  &
   $\Gamma(\rho \to \pi \pi)$ &
   $\Gamma(a_1 \to \pi \gamma)$ & $\Gamma(a_1 \to \rho \pi)$  \\
 $f_\pi$ & $\xi$ & $\zeta$ ($10^{-6}$) & $f_\rho$ & $f_{a_1}$ &
   $g_{\rho\pi\pi}$ & $r_\pi$(fm) & D/S ratio \\
\hline
expected &  &  &  $775.8\pm 0.5$ & $1230\pm 40$ & $146.4\pm 1.5$ &
 $0.640\pm 0.246$ & $250\sim 600$  \\
 $86.4\pm 9.7$ &  &  &  &  &  & $0.672\pm 0.008$ & $-0.108\pm 0.016$ \\
\hline
A & 3.125 & 0. & $[769.6]$ & $[1253]$ & 95.4 & 0. & 295.5 \\
85.0 & 4.0 & 0. & 138 & 163 & 4.8 & 0.585 & -0.055 \\
\hline
B & 2.836 & -5.930 & $[775.8]$ & $[1230]$ & $[146.5]$ & 0.088 & 165.3 \\
71.9 & 2.56 & -39.72 & 144 & 182 & 5.8 & 0.654 & $[-0.094]$ \\
\hline
C & 3.102 & -16.03 & $[775.8]$ & $[1246]$ & $[146.4]$ & 0.042 & 409.8 \\
$[78.7]$ & 4.010 & 0.09188 & 140 & 172 & 5.6 & 0.640 & -0.027 \\
\hline
\end{tabular}
\caption{\label{table1}
Various hadronic observables obtained in the present work.
The unit of masses, decay constants and decay widths is MeV.
The inputs for each case are shown in the brackets.}
\end{center}
}
\end{table}

The pion charge radius $r_\pi$ agrees with the experiment
in both case B and C.
Pion decay constant $f_\pi$ is a bit small in case B,
while the D$/$S ratio of $a_1 \to \rho \pi$ is small in case C,
compared with experiment.
As in other 5D models,
the KSRF relation $g_{\rho\pi\pi}^2/m_\rho^2=c/f_\pi^2$
with $c=1/2$~\cite{ksrf} is not satisfied very well.
In both case B and C, c is roughly $0.3$,
which means the complete vector meson dominance of
order ${\cal O}(p^2)$ four-pion interaction,
with the higher $\rho$ resonace and scalar exchange,
and contact four-pion interaction contribution below $\sim 10\%$.

\begin{table}[!htb]
\begin{center}
\small{
\begin{tabular}{|c|c|c|c|c|}
\hline
 &  case A & case B & case C & expected\\
\hline
$c_6^V$ & $0.$  & $0.0008$  & $0.0000$  &  $-0.0005$ \\
\hline
$c_6^A$ & $0.0014$ & $0.0000$  & $0.0006$  & $0.0008$ \\
\hline
$c_6$ & $-0.0014$ & $0.0008$  & $-0.0006$  & $-0.0013$ \\
\hline
\end{tabular}
\caption{\label{tableope}
OPE coefficients $c_6^V$, $c_6^A$, and $c_6$ in unit GeV$^6$.
}
}
\end{center}
\end{table}

The pion form factor $F(q^2)$ as a function of $q^2$
is plotted
in Figure~\ref{pionformfactor}.
We find the form factor has better behavior in case B and C
for large value of momentum than that in case A.

The OPE coefficients $c_6^V$, $c_6^A$, and $c_6$ are presented
in Table~\ref{tableope}.
The individual coefficients $c_6^V$ and $c_6^A$ do not agree very well
with the expected value in all three cases,
while the coefficient of left-right correlator
agrees with the expected value in case A.
However, we note that the OPE is also sensitive to the deformation of
the AdS metric~\cite{Hirn:2005vk}.
Considering the OPE behavior, it may be worth remarking that,
although the spirit of bottom-up AdS/QCD models has been to match
the theory in the UV and then compare with the physical observables
in the IR, it is not surprising that the best fit to data would arise
from a model that disagrees with the precise UV behavior of QCD,
where the model is not expected to be valid.


\section{Chiral Lagrangian for pseudoscalars up to $O(p^4)$}

Before we discuss the ${\cal O}(p^4)$ chiral Lagrangian,
we consider the vector field $\rho$ effective Lagrangian~\cite{Ecker:1989yg},
\begin{eqnarray}
{\cal L}_V &=& - \frac{1}{4} {\rm Tr}[V^{\mu\nu} V_{\mu\nu}]
  + \frac{1}{2} m_\rho^2 {\rm Tr} [V_\mu - \frac{i}{g} \Gamma_\mu]^2 \nonumber\\
&&  - \frac{1}{2\sqrt{2}} e g_{\gamma\rho} {\rm Tr} [V_{\mu\nu} f^{\mu\nu}_{+}]
  + \frac{i}{\sqrt{2}} f_{\rho\pi\pi} f_\pi^2 {\rm Tr} [V_{\mu\nu} u^\mu u^\nu]
\end{eqnarray}
with $\rho$ transforming as gauge field of SU$(2)_V$
and the notation of $\Gamma_\mu$, $f^{\mu\nu}_{+}$, $u^\mu$
the same as in ref.~\cite{Ecker:1989yg}.
The coefficients in the effective Lagrangian
are determined by matching with our theory with dim-6 operators.
The ${\cal O}(p^4)$ chiral Lagrangian for the pions
is given in ref~\cite{Ecker:1988te},
\begin{eqnarray}
{\cal L}_4 &=& L_1 {\rm Tr}^2 [ D_\mu U^\dagger D^\mu U ]
 + L_2 {\rm Tr} [ D_\mu U^\dagger D_\nu U ] {\rm Tr} [ D^\mu U^\dagger D^\nu U ]
  + L_3 {\rm Tr} [ D_\mu U^\dagger D^\mu U D_\nu U^\dagger D^\nu U ] \nonumber\\
&+&  L_4 {\rm Tr} [ D_\mu U^\dagger D^\mu U ]
   {\rm Tr} [ U^\dagger\chi +  \chi^\dagger U ]
  + L_5 {\rm Tr} [ D_\mu U^\dagger D^\mu U
   ( U^\dagger\chi + \chi^\dagger U )] \nonumber\\
&+& L_6 {\rm Tr}^2 [ U^\dagger\chi +  \chi^\dagger U ]
  + L_7 {\rm Tr}^2 [ U^\dagger\chi -  \chi^\dagger U ]
  +  L_8 {\rm Tr} [\chi^\dagger U \chi^\dagger U
  + U^\dagger\chi U^\dagger\chi ] \nonumber \\
&-& i L_9 {\rm Tr} [ F_R^{\mu\nu} D_\mu U D_\nu U^\dagger +
    F_L^{\mu\nu} D_\mu U^\dagger D_\nu U ]
  +  L_{10} {\rm Tr} [ U^\dagger F_R^{\mu\nu} U F_{L\mu\nu} ] \, .
\end{eqnarray}
In the present, we do not discuss scalar and pseudoscalar resonances contribution
to $L_{3,4,5,6,7,8}$,
and only study the vector and axial resonances contribution
to $L_{1,2,3,9,10}$.
After Integrating out the vector rho meson,
we obtain the following chiral coefficients,
\begin{eqnarray}
L_1 &=& \frac{f_\pi^4}{8 m_\rho^4} g_{\rho\pi\pi}^2 -
  \frac{f_\pi^4}{4 m_\rho^4} g_{\rho\pi\pi} f_{\rho\pi\pi},
 \qquad L_2 = 2 L_1, \qquad L_3 = - 6 L_1, \nonumber\\
L_9 &=& \frac{f_\pi^4}{m_\rho^4} g_{\rho\pi\pi}^2 +
  \frac{f_\pi^2}{2 m_\rho^2} e g_{\rho\pi\pi} f_{\rho\pi\pi} -
  \frac{2 f_\pi^4}{m_\rho^2} g_{\gamma\rho} g_{\rho\pi\pi} .
\end{eqnarray}
$L_{10}$ can be calculated from the two-point correlators
of vector and axial, $\Pi_{V,A}$,
\begin{eqnarray}
L_{10} &=& \frac{1}{4} [ \Pi^\prime_A(0) - \Pi^\prime_V(0) ],
\end{eqnarray}
where the derivative is over $p^2$.

We also calculate the electromagnetic mass difference of the pions
from the operator of ${\rm Tr}[Q_R U Q_L U^\dagger ]$,
\begin{eqnarray}
m_{\pi^+} - m_{\pi^0} & \simeq &
  \frac{3 \alpha_{\rm em}}{8 \pi m_\pi f_\pi^2}
   \int_0^\infty d p^2 (\Pi_A - \Pi_V) .
\end{eqnarray}


\begin{table}[!htb]
\begin{center}
\small{
\begin{tabular}{|c|c|c|c|c|c|c|}
\hline
case & $L_1$ & $L_2$ & $L_3$ & $L_9$ & $L_{10}$ & $m_{\pi^+}-m_{\pi^0}$ (MeV)\\
\hline
exp & $0.4\pm 0.3$ & $1.4\pm 0.3$ & $-3.5\pm 1.1$ & $6.9\pm 0.7$
 & $-5.5\pm 0.7$ & $4.6$ \\
\hline
A & $0.43$ & $0.86$ & $-2.6$ & $5.1$ & $-5.5$  & $3.4$ \\
\hline
B & $0.32$ & $0.65$ & $-1.9$ & $4.0$ & $-5.0$  & $1.5$ \\
\hline
C & $0.46$ & $0.93$ & $-2.8$ & $5.3$ & $-5.1$  & $2.9$ \\
\hline
\end{tabular}
\caption{\label{table2}
The chiral coefficients $L_i$ in unit $10^{-3}$.}
}
\end{center}
\end{table}
The chiral coefficients of relevance and
electromagnetic pion mass difference are given in Tab.~\ref{table2}.
Compared with Da Rold and Pomarol's case,
the results do not significantly change much in our two cases.

\section{Conclusions}

In this paper, we considered holographic QCD beyond the leading order,
by including two dim-6 dimension operators that go beyond
the usual quadratic kinetic terms for the bulk gauge field
$L_M$ and $R_M$, and scalar field $\Phi$ \cite{EKSS,DaRold:2005zs}.
We have studied the mass spectra, decay constants of
vector, axial and pseudoscalar sectors, and
phenomenology of $a_1 \to \rho \pi$, $\rho \to \pi \pi$
and $a_1 \to \pi \gamma$ channels.
In our work, we could achieve a non-vanishing branching ratio
for $a_1 \to \pi \gamma$, which is a new feature compared
with the usual holographic QCD in the leading order.
We also calculated the electromagnetic form factor of a charged pion,
(including the charge radius of a pion) which agrees with the experimental
results up to $q^2 \simeq 2$ GeV$^2$.
The numerical results are summarized in Table ~1, and compared with the
leading order results obtained by Da Rold and Pomarol \cite{DaRold:2005zs}
denoted as the case A.  We could achieve significant improvements in
overall phenomenology of the $\pi-\rho-a_1$ system
by including the $\kappa$ and $\zeta$ terms.

Let us remind ourselves that most studies based on the AdS/QCD approach
are just the leading order calculations, starting from the bulk Lagrangian
which is quadratic in the bulk gauge fields.  Including the next-to-leading
order corrections would be the next step to follow, and our present work makes
such an attempt by considering dim-6 operators that reduce to the $O( p^4 )$
operators after chiral symmetry breaking.  Considering the improvement
of overall phenomenology obtained in this work,
it would be clearly desirable to
have more systematic study of subleading corrections within AdS/QCD.

\acknowledgments

This work was supported in part by KOSEF SRC program through CHEP
at Kyungpook National University.

\end{document}